\documentclass[12pt,a4paper]{article}

\usepackage{amssymb,amsfonts,amsmath}
\usepackage{xcolor}

\title{Entanglement of observables:  Quantum conditional probability approach}

\author{Andrei Khrennikov and Irina Basieva\\
International Center for Mathematical Modeling\\
 in Physics and Cognitive Sciences \\
 Linnaeus University,  V\"axj\"o, Sweden}

\date{}

\begin{document}

\maketitle

\abstract{This paper is devoted to clarification of the notion of entanglement through decoupling it from the tensor product structure
and treating  as a constraint posed by probabilistic dependence of quantum observable $A$ and $B.$ In our framework, it is meaningless to speak about entanglement without pointing to the fixed observables $A$ and $B,$ so this is $AB$-entanglement. Dependence of quantum observables is formalized as non-coincidence of conditional probabilities.  Starting with this probabilistic definition, we achieve the Hilbert space characterization of the $AB$-entangled states as amplitude non-factorisable states.  In the tensor product case, 
$AB$-entanglement implies standard entanglement, but not vise verse. $AB$-entanglement for dichotomous observables is equivalent to their correlation, i.e.,  $\langle AB\rangle_\psi  \not=  \langle A\rangle_\psi \langle B\rangle_\psi .$ We describe the class of quantum states that are $A_u B_u$-entangled for a family of unitary operators $(u).$  Finally,  observables entanglement is compared with dependence of random variables in classical probability theory.}       
        
\section{Introduction}

We clarify and demystify the notion of entanglement by highlighting the observables role  and  exploring the quantum conditional probability calculus (see also \cite{JFA}). Observables' entanglement express the degree of dependence 
between two quantum observables $A$ and $B$ for a state  $|\psi\rangle.$ Hence,  it is meaningless to speak about entanglement without selecting the fixed observables $A$ and $B.$ In this aspect our work is close to the papers \cite{Z1,Z2}; we cite \cite{Z2}: 	

{\it ``Here we propose that a partitioning of a given Hilbert space is induced by the experimentally accessible observables .... In this sense entanglement is always relative to a particular set of experimental capabilities.''} 

The distinguishing feature of our work is exploring the calculus of quantum conditional probabilities (see \cite{JFA} for the first step in this direction). 

Observables' entanglement differs from the standard tensor product non-factorization 
approach, see, e.g., \cite{WER}: 

{\it ``A state is said to be entangled if it cannot be written as a convex sum of tensor product states.''} 

In this paper we work only with pure states. 

 \subsection{Entanglement as Hilbert space representation of observables dependence}

Dependence of observables is formalized on the basis of the calculus of quantum conditional probabilities \cite{Koopman,BL,BL1,KHRWS}. For dichotomous observables $A$ and $B,$ a state $|\psi\rangle$ is $AB$-entangled   iff
\begin{equation}
\label{li}
P(B=+| A=+; \psi) \not=  P(B=+| A=-; \psi),
\end{equation}
where conditional probabilities are defined via the combination of the Born rule and the L\"uders projection postulate \cite{Luders}.

There is nothing mystical in this notion of $AB$-entanglement. The difference between conditional probabilities quantifies the degree of dependence of $B$-outcome on $A$-outcomes. This measure of $AB$-entanglement can be expressed in the Hilbert space terms and it is similar to the concurrence-measure for the usual entanglement. 

$AB$-entanglement is not coupled to the tensor product structure. It can be considered e.g. for a state space 
${\cal H}$ having non-factorisable dimension, say $\rm{dim} \; {\cal H}= 5.$ Generally $AB-$entanglement is introduced for 
non-commuting operators (incompatible observables). But, in the main part of this paper we work with  commuting 
operators (with dichotomous spectrum). As can be expected, for non-commuting operators $AB$- and $BA$-entanglements do not coincide. 

The main part of the article is devoted to the Hilbert space characterization of $AB$-entangled quantum states. The standard entangled states are defined as non-factorisable w.r.t. the tensor product structure on  ${\cal H}.$ We demonstrate that, for compatible dichotomous observables, $AB$-entangled states can be characterized as amplitude non-factorisable states. Here amplitudes are defined as norms of the state projections on subspaces ${\cal H}_{AB}(\alpha \beta)$ consisting of common eigenvectors of the operators (with eigenvalues $\alpha, \beta).$ Such expansion w.r.t. to the joint eigenvectors was actively explored in works of Ozawa, e.g., 
\cite{Ozawa2009}, devoted to a similar problem.

Conditional probability treatment of entanglement can be considered as indirect contribution to critique of nonlocal interpretation of quantum mechanics (QM) and connection of the violation of the Bell inequalities \cite{Bell0}-\cite{CHSH} with spooky action at a distance (see, e.g.,  \cite{Loub}--\cite{HP} for analysis of the nonlocality problem within the QM-formalism; see also for probabilistic analysis \cite{Kupczynski1}-\cite{Boughn1a}). 

\subsection{Observables entanglement and correlation}

The description of entangled states as amplitude non-factorisable states leads to the following result. For compatible dichotomous observables, a quantum state $|\psi\rangle$ is  $AB$-entanglement iff correlation of these observables w.r.t. this state is not factorisable, i.e.,  
\begin{equation}
\label{li1}
\langle AB\rangle_\psi  \not=  \langle A\rangle_\psi \langle B\rangle_\psi ,    
\end{equation}
or observables covariance is non-zero:
\begin{equation}
\label{li1a}
\rm{cov}\; (A,B; \psi) \equiv \langle (A - \langle A\rangle_\psi) (B-\langle B\rangle_\psi) \rangle_\psi  = \langle AB\rangle_\psi  - \langle A\rangle_\psi \langle B\rangle_\psi \not=0,    
\end{equation} 

Let ${\cal H}={\cal H}_1 \otimes {\cal H}_2.$ In this case, it is worth to compare $AB$-entanglement and usual entanglement w.r.t. to the correlation property. If a quantum state $|\psi\rangle$ is factorisable, i.e.,
$|\psi\rangle= |\phi_1\rangle\otimes |\phi_2\rangle,$ then, of course, correlation is factorisable for any pair of operators, 
\begin{equation}
\label{li2}
\langle AB\rangle_\psi  =  \langle A\rangle_\psi \langle B\rangle_\psi, \; \mbox{i.e.} \;     \rm{cov}\; (A,B; \psi) =0.
\end{equation}
 However, the inverse statement is not correct. For some entangled states, the equality (\ref{li2}) holds true. The latter is impossible for $AB$-entangled states.

\subsection{Observables entanglement vs. random variables dependence in classical probability}

We recall that in classical probability theory independence of random variables $A$ and $B$ implies correlation factorization 
\begin{equation}
\label{li3}
\langle AB \rangle_{{\cal P}}  = E_{{\cal P}} [AB]= 
E_{{\cal P}}[A]E_{{\cal P}}[B]=  \langle A \rangle_{{\cal P}}   \langle B \rangle_{{\cal P}} ,  \; \mbox{i.e.} \;     \rm{cov}\; (A, B; {\cal P}) =0
\end{equation}
where $E$ denotes the mathematical expectation for a classical probability space  ${\cal P}$ 
(see appendix on the notion of Kolmogorov probability space \cite{K}).

).  In textbooks 
on classical probability theory one proceeds with a fixed probability space  ${\cal P},$ so dependence of all quantities on 
${\cal P}$ is not visualized.

For  classical dichotomous random variables, independence is equivalent  to uncorrelation; hence, for such random variables dependence is equivalent to correlation, i.e., non-factorization of product's average. We obtain the same characterization of compatible dichotomous $AB$-entangled quantum observables. For such observables, the notion of observables entanglement is simply the Hilbert space representation of the notion of dependence for classical random variables, $\rm{cov}\; (A,B; {\cal P}) \not=0.$
Generally observables entanglement extends the notion of dependence of random variables, within the Hilbert space formalism. 

In section \ref{EX}, we study $A_u B_u$-entanglement for families of dichotomous pairwise compatible observables, i.e., represented by commuting operators, $[A_u, B_u]=0,$  ${\cal U}=(u)$ is a family of unitary  transformations. Here generally $[A_u, A_v]\not=0, [B_u,B_v]\not=0.$ And we found an analog of the singlet state $|\psi_{{\cal U}}\rangle,$ i.e., a state that is $(A_u,B_u)$-entangled for any  transformation $u \in {\cal U}.$

This construction highlights the essence of quantum entanglement, as formalized within quantum probability calculus in complex Hilbert space. For each concrete pair of compatible dichotomous  observables, $A_u B_u$-entanglement is just the Hilbert space representation of dependence of classical random variables or equivalently (since this is the dichotomous case) correlation of random variables. So for each pair, there exists a classical probability space ${\cal P}_u$ such that    $\rm{cov}\; (A_u, B_u; {\cal P}_u) \not=0,$ i.e., correlation of these two observables can be described within classical probability theory. However, generally it is impossible to construct a 
probability space ${\cal P}$ which would describe dependence for all pairs of observables $A_u, B_u.$ In contrast to this 
classical impossibility, quantum probability can describe all these correlations, as correlations w.r.t. the state $|\psi_{{\cal U}}\rangle.$

This example illustrates the general advantage of quantum entanglement in the description of observables' correlations. Thus, the nonclassical essence of quantum observables entanglement is the possibility to model correlations for a wider class of observables than in classical probability theory. Hence, our approach to entanglement highlights not the strength of correlations, but the possibility to realize a wide class of correlated observables for the same quantum state (preparation procedure \cite{Muynck}). 

\subsection{EPR-argument and Schr\"odinger's viewpoint on entanglement}

Although this article is mainly directed to the mathematical analysis of the probabilistic structure of QM, in the form of complex HIlbert space representation, we would like to make the following foundational remark.

The probabilistic viewpoint on the ``EPR-paradox'' \cite{EPR} is presented in  Schr\"odinger's  paper  \cite{SCHE,SCHE1} that initiated the modern theory of entanglement. However, this theory ignores the important message of Schr\"odinger: entanglement characterizes probability update for the outcomes  of observable $B$ conditioned on the  outcomes of observable $A.$ In the framework of \cite{SCHE,SCHE1}
it is meaningless to speak about entanglement  without specifying the observables. The state update -- the Hilbert space representation of the probability update -- encodes the procedure of conditional prediction. For Schr\"odinger, the quantum formalism is a mathematical machinery for probability prediction and a quantum state is a part  of such machinery (cf. with the V\"axj\"o interpretation 
of QM \cite{Vaxjo2004,KHR5} and QBism \cite{Fuchs6}  as well as with reality without realism interpretation of QM \cite{PL,PLR}).    
We can say that Schr\"odinger interpreted
quantum probabilities  as conditional probabilities. This viewpoint was latter  expressed in the works of 
 Koopman \cite{Koopman}, Ballentine \cite{BL, BL1}, Khrennikov \cite{KHRWS}.

From the probabilistic viewpoint the EPR argument \cite{EPR} is based outcomes' prediction with  conditional probability one. In the Hilbert space formalism such update is represented with the special class of observables entangled states (see \cite{JFA} for EPR entangled states; see also Ozawa \cite{Ozawa2005}).

\section{Quantum  conditional probability}

All probabilities under consideration depend of quantum state $|\psi\rangle,$ i.e., $P\equiv P_\psi,$ but to make the notation shorter, we omit the state index. 

In the quantum formalism an observable $A$ (with a discrete range of values) is represented by a Hermitian operator 
\begin{equation}
\label{L2a}
 A= \sum_\alpha \alpha  E_A(\alpha),
\end{equation} 
where $ E_A(\alpha)$ is projection onto the space ${\cal H}_A(\alpha)$ of eigenvectors for the eigenvalue $\alpha.$ For a pure state $\vert \psi \rangle,$ the probability to get the outcome $A=\alpha$ is given by the Born's rule: 
\begin{equation}
\label{L2a1}
p(A=\alpha)= \Vert   E_A(\alpha) \psi \Vert^2. 
\end{equation} 
A measurement with the  outcome $A= x$ generates back-action onto system's state:
\begin{equation}
\label{L2b}
\vert \psi \rangle \to \vert \psi \rangle_\alpha^A=   E_A(\alpha)\vert \psi \rangle/ \Vert  E_A(\alpha) \psi \Vert.
\end{equation} 
This is the projection postulate in the L\"uders form \cite{Luders}. We remark that von Neumann \cite{VN} applied this postulate only to observables with non-degenerate spectra. For observables with degenerate spectra, he considered more general class of back-actions. Nowadays, this class if formalized within theory of quantum instruments \cite{DV,Oz1}.

The projection postulate is the mathematical tool for quantum conditioning. Measurements of another observable, say $B,$ conditioned on 
the outcome $A=\alpha$ lead to probability (see, e.g., \cite{KHRWS}): 
\begin{equation}
\label{Lx3m}
P(B=  \beta| A= \alpha, \psi)= \Vert  E_B(\beta)  \psi_A^{\alpha}\Vert^2=
\frac{ \Vert  E_B(\beta)   E_A(\alpha) \psi \Vert^2}{\Vert E_A(\alpha)  \psi \Vert^2}.
\end{equation}
We shall use this definition of conditional probability to define the special form of entanglement.  
    		
\section{Observables entanglement}
\label{ENTAB}

Consider two (generally incompatible) observables $A$ and $B$  represented by Hermitian operators (denoted by the same symbols) with eigenvalues  $X_a= \{\alpha_1,..., \alpha_m\}$ and $X_b= \{\beta_1,..., \beta_k\}.$ 

{\bf Definition 1.}    {\it In the state $|\psi\rangle,$ the outcome $B=\beta$ depends on the outcomes of $A$ if for at least two values of $A, \alpha= \alpha_i, \alpha_j,$
the corresponding conditional probabilities don't coincide,} 
\begin{equation}
\label{G1}
P( B= \beta| A= \alpha_i) \not= P( B= \beta| A= \alpha_j),
\end{equation} 

Thus,  the probability to get the outcome  $B=\beta$ if  the preceding $A$-measurement had the outcome $A=\alpha_i$ differs from 
the probability to get the same outcome $B=\beta$ if  the preceding $A$-measurement had the outcome $A=\alpha_j.$ 

The outcome $B=\beta$ does not depend on the outcomes of $A$ iff
\begin{equation}
\label{G1q}
P( B= \beta| A= \alpha_i) = P( B= \beta| A= \alpha_j),\; \mbox{for all pairs}  \;\alpha_i, \alpha_j,
\end{equation} 
i.e. the conditional probability for this outcome is constant w.r.t. the outcomes of $A.$ Denote it $P( B= \beta| A).$
Does it coincide with unconditional probability $P( B= \beta)?$ (see section \ref{dep}).

{\bf Definition 2.}    {\it  For observables $A$ and $B,$  a state $|\psi\rangle$ is called $AB$-entangled if all outcomes of $B$ depend on outcomes of $A,$ i.e. for all $\beta$ condition (\ref{G1}) holds  for some  $ \alpha_i, \alpha_j.$ }

\medskip

We note that this is the purely  probabilistic definition, it does not involve the notion of Hilbert space, it can be applied to any statistical physical theory in that system's state generates conditional probabilities. This definition formalizes dependence of observables. In section \ref{dep} we present the detailed comparison with the notion of dependence of random variables in classical probability theory. For classical random variables, $AB$-entanglement due to Definition 2 is reduced to the well known notion of 
dependent random variables. 

For incompatible observables, $AB$-entanglement does not imply  $BA$-entanglement. A state $|\psi\rangle$ is called  $A\leftrightarrow B$-entangled if it is both $AB$- and $BA$-entangled. For compatible observables,  $AB$- and $BA$-entanglements w.r.t. to some state
are equivalent (see section \ref{dep}). 

We remark that condition  (\ref{G1}) characterizing $AB$-entanglement should be completed by the conditional probability existence 
constraints:
\begin{equation}
\label{G2m}
 P(A= \alpha)\not= 0, \;  \alpha \in X_a .
\end{equation} 

In the Hilbert space formalism the basic inequality (\ref{G1}) determining outcome dependence can be written as 
 \begin{equation}
\label{G20}
\frac{ \Vert  E_B(\beta)   E_A(\alpha_i) \psi \Vert^2}{\Vert E_A(\alpha_i) \psi \Vert^2}\not= 
\frac{ \Vert  E_B(\beta)   E_A(\alpha_j) \psi \Vert^2}{\Vert E_A(\alpha_j) \psi \Vert^2}.
\end{equation}
 It implies that 
\begin{equation}
\label{G2j1}
|| E_B(\beta) E_A(\alpha_i)  \psi||\; || E_A(\alpha_j)  \psi||\not= 
|| E_B(\beta) E_A(\alpha_j)  \psi||\; || E_A(\alpha_i)  \psi||.
\end{equation} 

{\bf Proposition 1.} {\it Inequality  (\ref{G2j1}) is necessary and sufficient condition for  dependence of the outcome $B=\beta$ on the outcomes $A=\alpha_i, \alpha_j$.}

{\bf Proof.}  1). Inequality (\ref{G1}) implies inequality (\ref{G2j1}).

2). Now let inequality  (\ref{G2j1}) hold for some $\beta$ and $\alpha_i, \alpha_j.$ 

Generally we have 
$$
P(A=\alpha_i)= ||E_A(\alpha_i) \psi||^2  = ||\sum_\beta   E_B(\beta)E_A(\alpha_i) \psi||^2 =  
\sum_\beta  || E_B(\beta)E_A(\alpha_i) \psi||^2.
$$ 
 Hence, if $P(A=\alpha_i)=0,$ then, for any $\beta,$   $|| E_B(\beta)E_A(\alpha_i) \psi||^2=0.$ Hence,  inequality (\ref{G2j1}) is violated. Thus, inequality (\ref{G2j1}) implies that $P(A=\alpha_i)\not=0$ and  $P(A=\alpha_j)\not=0;$  conditional probabilities are well defined. Hence, 
inequality (\ref{G2j1}) can be written in the form (\ref{G1}). So, the outcome $B=\beta$  depends on the outcomes 
$A=\alpha_i, \alpha_j.$ 

{\bf Proposition 1a.} {\it Let observable $A$ be dichotomous.  Then inequalities  
\begin{equation}
\label{G2p}
||  E_B(\beta) E_A(+) \psi || \;||E_A(-) ||  \not= || E_B(\beta) E_A(-) \psi || \; ||E_A(+)||, \; \; \beta \in X_b, 
\end{equation} 
are the necessary and sufficient conditions for   $AB$-entanglement.}
\medskip

{\bf Proposition 2.} {\it $AB$-entanglement is invariant under unitary  transformations.}

{\bf Proof.} Let $U$ be a unitary operator and let $A_U=UAU^\star, B_U=UBU^\star, \vert \psi_U \rangle= U  \vert \psi \rangle.$
Then   $
|| E_{B_U}(\beta) E_{A_U}(\alpha) \psi_U \rangle =  || E_{B}(\beta) E_{A}(\alpha) \psi||
$
and  $|| E_{A_U}(\alpha)  \psi_U||= || E_{A}(\alpha_i) \psi ||.$

\subsection{Measure of observables' entanglement}

Now we introduce a quantitative measure of $AB$-entanglement. 

{\bf Definition 3.} {\it   For quantum observables $A$ and $B,$ $AB$-concurrence  in the state $|\psi\rangle$ is defined as}
\begin{equation}
\label{ME}
{\cal M}_{AB}(\psi)= \sum_\beta\sum_{\alpha \not=\alpha^\prime} 
|P( B= \beta| A= \alpha) - P( B= \beta| A= \alpha^\prime)|.
\end{equation} 
(see section \ref{4d} on comparison with the standard concurrence measure of entanglement). 
The crucial issue is that $AB$-concurrence depends on a pair of observables. The maximally $AB$-conditional probability entangled states are  considered in section \ref{EPRE}, these are the EPR-states invented in \cite{JFA}.

In operator terms, $AB$-concurrence in state $|\psi\rangle$ is given by 
\begin{equation}
\label{MEOa}
{\cal M}_{AB}(\psi)= \sum_\beta\sum_{\alpha \not=\alpha^\prime} 
\Big| \; \frac{||E_B(\beta) E_A(\alpha) \psi||^2}{||E_A(\alpha) \psi||^2}- 
\frac{||E_B(\beta) E_A(\alpha^\prime) \psi||^2}{||E_A(\alpha^\prime)\psi||^2} \; \Big|.
\end{equation} 
We also define re-normalized $AB$-concurrence as
\begin{equation}
\label{MEO}
{\cal C}_{AB}(\psi)= \sum_\beta \sum_{\alpha \not=\alpha^\prime} 
\Big| \; ||E_B(\beta) E_A(\alpha) \psi||\; ||E_A(\alpha^\prime) \psi|| - 
||E_B(\beta) E_A(\alpha^\prime) \psi|| \; ||E_A(\alpha)\psi|| \; \Big|.
\end{equation}

\subsection{Dichotomous observables: equivalence of outcomes' dependence and entanglement}

{\bf Proposition 3.} {\it For dichotomous  observables $A$ and $B,$  dependencies  of
 the values  $B=-$ and $B=+$  on the outcomes of $A$ are equivalent. Thus, each dependence is equivalent to $AB$-entanglement.}   
\medskip

{\bf Proof.} In the state $|\psi\rangle$ the value $B=-$ depends on the outcomes of $A$ if 
\begin{equation}
\label{G3}
P( B= -| A= +) \not= P( B=  - | A= -),
\end{equation} 
This automatically implies that even the value $B=+$ depends on the outcomes of $A,$
$$
P( B= +| A= +) = 1- P( B= -| A= +)   \not=  
$$
$$
1- P( B= -| A= -)  = P( B=  + | A= -),
$$
i.e., the state $|\psi\rangle$ is $AB$-entangled. 

{\bf Proposition 3a.} {\it For dichotomous  observables $A$ and $B,$ $AB$-entanglement is characterized by the single constraint:}
\begin{equation}
\label{G2pb}
||  E_B(+) E_A(+) \psi || \;||E_A(-) ||  \not= || E_B(+) E_A(-) \psi || \; ||E_A(+)||, 
\end{equation} 

A state $|\psi\rangle$ is $AB$-disentangled, iff 
\begin{equation}
\label{G2pa}
||  E_B(+) E_A(+) \psi || \;||E_A(-) || = || E_B(+) E_A(-) \psi || \; ||E_A(+)||.
\end{equation} 
 
\subsection{EPR-entanglement}
\label{EPRE0}

In article \cite{JFA} we invented the notion of entanglement which is also coupled to the probability update via conditional measurement, so called  {\it EPR-entanglement.} Consider two observables, generally incompatible, represented by operators $A$ and $B.$

{\bf Definition 1a.}  {\it In the state $|\psi\rangle,$ the observables $A$ and $B$ are perfectly  conditionally correlated (PCC) for the values  $(A=\alpha, B=\beta)$ if the conditional probability to get the outcome  $B=\beta$ if  the preceding $A$-measurement had the outcome $A=\alpha$ equals to 1,}
\begin{equation}
\label{Lx3}
P(B=  \beta| A= \alpha)= \frac{ \Vert E_B(\beta)   E_A(\alpha) \vert \psi \rangle \Vert^2}{\Vert E_A(\alpha) \vert \psi \rangle\Vert^2} =  1.
\end{equation}
The order of observations is important even for compatible observables and to be more precise we have to speak about $AB$-PCC.  
More generally consider observables with values $(\alpha_i)$ and $(\beta_i)$ and some set $\Gamma$ of pairs $(\alpha_i, \beta_j).$   

\medskip

{\bf Definition 2a.}  (EPR entanglement) {\it If, for all pairs from the set $\Gamma,$   $\vert \psi \rangle$ is PCC-state, then such state is called EPR entangled w.r.t. $\Gamma.$}

\medskip

 We are interested in sets $\Gamma$ such that each of $\alpha$ and $\beta$ values appears in the pairs once and only once.  We call such {\it EPR entanglement complete.}

For example, for two dichotomous observables with $\alpha, \beta=\pm 1,$ 
we consider, e.g.,  the set of the  pairs $(A=+, B=-), (A=-, B=+),$ in short, EPR $A=-B$ entanglement, 
or the  pairs $(A=+, B=+), (A=-, B=-),$ EPR $A=B$ entanglement. Consider such EPR-entanglements. 

Let us start with $A=-B$ entanglement, i.e., 
$P(B=-|A=+)=1$ and $P(B=+|A=-)=1.$ Thus, $P(B=+|A=+)=0$ and $P(B=-|A=-)=0,$ and
$P(B=-|A=+)=1 \not= P(B=-|A=-)=0$ and    $P(B=+|A=-)=1 \not= P(B=+|A=+)=0.$ In this case EPR-entangled state is automatically
$AB$-entangled  in the sense of the present paper. 

In the same way $A=B$ entanglement, i.e., 
$P(B=+|A=+)=1$ and $P(B=-|A=-)=1.$ Thus, $P(B=-|A=+)=0$ and $P(B=+|A=-)=0,$ and
$P(B=+|A=+)=1 \not= P(B=+|A=-)=0$ and    $P(B=-|A=-)=1 \not= P(B= -|A=+)=0.$ And again EPR-entangled state is automatically $AB$-entangled.

Thus, EPR $AB$-entanglement is just the very special case of $AB$-entanglement.   
		
\subsection{Maximally $AB$-entangled states}
\label{EPRE}

For dichotomous observables the measure of $AB$-entanglement (\ref{ME}) has the form 
$$
{\cal M}_{AB}(\psi)= |P( B=+| A= -) - P( B= +| A= +)|
$$
$$+  |P( B=-| A= -) - P( B= -| A= +)|,
$$
hence it can writen as 
\begin{equation}
\label{G4ee}
{\cal M}_{AB}(\psi) = 2|P( B=+| A= -) - P( B= +| A= +)|. 
\end{equation}
From this formula, we immediately obtain the following characterization of maximally $AB$-entangled states: 

{\bf Proposition 4.} {\it The $AB$-concurrence ${\cal M}_{AB}$ approaches its maximal value, ${\cal M}_{AB}(\psi)=2,$ if and only if $|\psi\rangle$ is EPR $AB$-entangled.} 

\section{Joint eigenvectors' superposition }
\label{eigen}

Our aim is find interrelation between $AB$-entanglement and the standard one. The latter is considered in the case of compatible  observables.\footnote{ And, as was stressed, typically one does not point out to the role of observables;  the notion of  entanglement is  associated solely with a state. In our approach the pair of observables $A$ and $B$ acts as the basic counterpart of the notion of $AB$-entanglement.} 

Now we assume, except section \ref{dep} that observables are dichotomous and compatible, i.e.,  represented by commuting Hermitian operators $A$ and $B$, $[A,B]=0,$ with eigenvalues  $\alpha, \beta= \pm 1.$ 

 The state space can be represented as the direct sum of subspaces for the joint eigenvectors, $ A \vert \phi_{\alpha \beta}\rangle=\alpha |\phi_{\alpha \beta}\rangle,  B \vert \phi_{\alpha \beta}\rangle=  \beta |\phi_{\alpha \beta}\rangle,\alpha, \beta = \pm,$   
\begin{equation}
\label{En1}
{\cal H} = {\cal H}_{ab}(++) \oplus {\cal H}_{ab}(+-)\oplus {\cal H}_{ab}(-+)  \oplus {\cal H}_{ab}(--).
\end{equation}  

Any (normalized) vector can be represented as superposition
\begin{equation}
\label{E2}
\vert \psi\rangle = \vert \psi_{++} \rangle +\vert \psi_{+-}\rangle  + \vert \psi_{-+}\rangle + \vert \psi_{--}\rangle,
\end{equation}
 where  $\vert \psi_{\alpha \beta}\rangle \in {\cal H}_{ab}(\alpha \beta)$ and 
\begin{equation}
\label{E3}
1= \Vert  \psi \Vert^2 = \Vert \psi_{++} \Vert^2 + \Vert  \psi_{+-}\Vert^2  + 
\Vert  \psi_{-+} \Vert^2 + \Vert  \psi_{--} \Vert^2.
\end{equation}
To be able to determine the $A$-conditional probabilities, we have to assume that      
\begin{equation}
\label{E4}
\vert \psi_{++}\rangle + \vert \psi_{+-}\rangle \not=0, \; \vert \psi_{-+}\rangle + \vert \psi_{--}\rangle \not=0.
\end{equation}
This is the constraint for operating with  $AB$-entanglement. 

We calculate the quantities for  formula (\ref{G2p})
$$
\langle  \psi|E_A(+)| \psi\rangle= \Vert \psi_{++} \Vert^2 + \Vert  \psi_{+-} \Vert^2, \; 
\langle  \psi|E_A(-)| \psi\rangle =  \Vert \psi_{-+}\Vert^2 + \Vert   \psi_{--} \Vert^2.
$$
$$
\langle  \psi|E_B(-) E_A(+)| \psi\rangle=  \Vert  \vert \psi_{+-}\rangle\Vert^2, \; 
\langle  \psi|E_B(-) E_A(-)| \psi\rangle=  \Vert  \vert \psi_{--}\rangle\Vert^2.
$$
Inequality (\ref{G2j1}) has the form:
$$
\Vert  \psi_{+-}\Vert^2 \Big(\Vert  \psi_{-+}\Vert^2 + \Vert  \psi_{--}\Vert^2\Big) \not= 
$$
$$
\Vert  \psi_{--}\Vert^2 \Big( \Vert \psi_{++}\Vert^2 + \Vert   \psi_{+-}\Vert^2\Big)
$$
And it is reduced to the inequality
\begin{equation}
\label{E5}
\Vert  \psi_{+-} \Vert^2 \Vert  \psi_{-+} \Vert^2 \not= \Vert  \psi_{--}\Vert^2 \Vert  \psi_{++}\Vert^2.
\end{equation}

By Proposition 1 this is the necessary and sufficient condition for $AB$-entanglement. We note that the constraint (\ref{E4}) that is needed to be able to define quantum conditional probabilities is automatically encoded in (\ref{E4}).

\section{Tensor product case}
\label{TPS}

Now let us equip the state space with a tensor product structure 
${\cal H}=   {\cal H}_1 \otimes   {\cal H}_2$ and represent the operators $A$ and $B$ as tensor products, i.e.,
$A= a \otimes I$ and $B= I \otimes b,$ where  $a:   {\cal H}_1 \to   {\cal H}_1$ and $b:   {\cal H}_2 \to 
  {\cal H}_2.$ We remark that $  {\cal H}_1=     {\cal H}_1(+) \oplus   {\cal H}_1 (-)$ and    $  {\cal H}_2=     {\cal H}_2(+) \oplus   {\cal H}_2(-).$ Each $|\phi_A \rangle \in   {\cal H}_1$ and   $|\phi_B \rangle \in   {\cal H}_2$ can be represented as superpositions:
$|\phi_A \rangle= |\phi_{a,-} \rangle + |\phi_{a.+} \rangle$ and $|\phi_B \rangle= |\phi_{b,-} \rangle + |\phi_{b.+} \rangle.$ Hence, 
for a factorisable state  $|\psi \rangle= |\phi_A \rangle |\phi_B \rangle,$ superposition (\ref{E2}) has the form:
\begin{equation}
\label{E6}
|\psi \rangle= |\phi_{a,+} \rangle |\phi_{b,+} \rangle + |\phi_{a,+} \rangle |\phi_{b,-} \rangle + 
|\phi_{a,-} \rangle |\phi_{b,+} \rangle + |\phi_{a,-} \rangle |\phi_{b,-} \rangle
\end{equation}    
Hence, 
\begin{equation}
\label{E7}
\Vert  \psi_{+-} \Vert^2 \Vert  \psi_{-+} \Vert^2 = \Vert  \psi_{--}\Vert^2 \Vert  \psi_{++}\Vert^2.
\end{equation}
So, we proved the following simple, but important theorem: 

\medskip

{\bf Theorem 1.} {\it Let ${\cal H}=   {\cal H}_1 \otimes   {\cal H}_2$ and $A= a \otimes I, B= I \otimes b,$ where operators $a, b$ have dichotomous spectra. Then any $AB$-entangled state $|\psi \rangle$ is also entangled w.r.t. the tensor product structure.}

\medskip

Hence,   $AB$-entangled states form the subclass ${\cal E}_{AB}$ of the class of all entangled states ${\cal E}$ for the tensor product decomposition ${\cal H}=   {\cal H}_1 \otimes   {\cal H}_2.$  We shall see that ${\cal E}_{AB}$ is a proper subclass of ${\cal E}.$

\section{$AB$-entanglement as amplitude factorization and  observables' correlation}

The standard entanglement is characterized via the condition of state non-factorization w.r.t. the tensor product structure. We want to find an analogous condition for $AB$-entanglement. We start with studying the case of four dimensional state space. 

\subsection{Four dimensional state space}
\label{4d}

Consider now two commuting operators $A$ and $B$ acting in four dimensional space ${\cal H}$ having dichotomous spectra $\alpha, \beta = \pm 1$ and common eigenvectors for all possible combinations of $\alpha, \beta,$ $A \vert \alpha \beta \rangle = \alpha \vert \alpha \beta \rangle, B \vert \alpha \beta \rangle = \beta \vert \alpha \beta \rangle,$ i.e., all joint eigensubspaces are one dimensional.
In this case ${\cal H}$ and the operators can be represented in the tensor form, ${\cal H}= {\cal H}_1 \otimes {\cal H}_2$ (two qubit state space) and 
$A= a \otimes I, B=I \otimes b;$ we can use the standard notion of entanglement and compare it with $AB$-entanglement.  

Any state can be represented as superposition
\begin{equation}
\label{E2b}
\vert \psi\rangle = c_{++}\vert ++ \rangle + c_{+-} \vert +-\rangle  + c_{-+}\vert -+\rangle + c_{--}\vert --\rangle,
\end{equation}
 where  $|c_{++}|^2 + |c_{+-}|^2 + |c_{-+}|^2 + |c_{--}|^2 =1.$ 

This state is factorisable if and only if the coefficients satisfy  the following equality:
\begin{equation}
\label{Em2}
 c_{+-} c_{-+} = c_{--} c_{++}.
\end{equation}

One of the measures of standard entanglement is concurrence and, for  $\vert \psi\rangle$,  concurrence is given by 
\begin{equation}
\label{Em2c}
{\cal C}(\psi)= |c_{+-} c_{-+} - c_{--} c_{++} |.
\end{equation}
So. a state is entangled iff ${\cal C}(\psi)\not= 0.$
Re-normalized $AB$-concurrence (\ref{MEO}) is written as
\begin{equation}
\label{Em2kc}
{\cal C}_{AB}(\psi)= |\;|c_{+-} c_{-+}| - |c_{--} c_{++}|\; |.
\end{equation}
So. a state is $AB$ entangled iff ${\cal C}_{AB}(\psi)\not= 0.$

\medskip

{\bf Definition 4.} {\it A state $|\psi \rangle$ is called $AB$ amplitude-factorisable w.r.t. the pair of compatible observables $A$ and $B$ with dichotomous spectrum 
if the absolute values of the coordinates in the joint eigenvectors basis are factorisable, i.e., for (\ref{E2b}), }
\begin{equation}
\label{Ea1ttt}
|c_{\alpha \beta}|= \lambda_\alpha\;  \mu_\beta, \; \mbox{where}  \; \alpha, \beta= \pm,   
\end{equation}
{\it where $\lambda_\alpha, \;\mu_\beta \geq 0$ and }
\begin{equation}
\label{Ea2b}
\lambda_+^2  + \lambda_-^2 =1, \; \mu_+^2 + \mu_-^2= 1.
\end{equation}

\medskip

The state factorization implies the amplitude  factorization, but not vice verse: condition (\ref{Ea1ttt}) does not take into account 
the presence of phases, but the latter can generate the standard entanglement even for amplitude-factorisable states.

Let $$
|\psi \rangle=  (\lambda_+ |+ \rangle + e^{i \theta_1} \lambda_- |- \rangle)  \otimes 
(\mu_+ |+ \rangle + e^{i \theta_2} \mu_- |- \rangle) =
$$
$$
\lambda_+ \mu_+ |++ \rangle +  e^{i \theta_2} \lambda_+\mu_- |+ -\rangle + e^{i \theta_1} \lambda_- \mu_+ |-+ \rangle + e^{i (\theta_1+\theta_2)} \lambda_- \mu_- |-- \rangle
$$
Now consider the state 
$$
|\psi \rangle =|++ \rangle +  |+ -\rangle -  |-+ \rangle +|-- \rangle,
$$
it is entangled in ordinary sense, although it is amplitude factorisable.

The violation of relation (\ref{E5}) has the form of the equality:
\begin{equation}
\label{E5b}
|c_{+-}|^2 |c_{-+}|^2 = |c_{--}|^2 |c_{++}|^2
\end{equation}

{\bf Theorem 2.} {\it Let $\rm{dim}\; {\cal H}= 4.$ For a pair of compatible dichotomous observables $A$ and $B,$ a state $|\psi \rangle$ is $AB$-entangled if and only if it is amplitude non-factorisable.}

{\bf Proof.} As we have seen, $AB$-disentangled is characterized by the equality (\ref{E5b}). We prove that this equality is equivalent to the amplitude factorization. It is clear that amplitude factorization implies (\ref{E5b}). Now let this equality holds true. We set   (under the condition that the denominators are not equal to zero): 
\begin{equation}
\label{PR1}
\lambda_+=\frac{|c_{++}|}{\sqrt{|c_{++}|^2 + |c_{-+}|^2}},  \lambda_-=\frac{|c_{-+}|}{\sqrt{|c_{++}|^2 + |c_{-+}|^2}}.
\end{equation} 
\begin{equation}
\label{PR2}
\mu_+= \frac{|c_{++}|}{\sqrt{|c_{++}|^2 + |c_{+-}|^2}},
\mu_-=\frac{|c_{+-}|}{\sqrt{|c_{++}|^2 + |c_{+-}|^2}}.
\end{equation} 
It is easy to check that these quantities factorize the modules of the coordinates $c_{\alpha, \beta}.$ For example, 
$$
\lambda_+ \mu_- = |c_{++}| |c_{+-}|/ \sqrt{(|c_{++}|^2 + |c_{-+}|^2)(|c_{++}|^2 + |c_{+-}|^2)}.
$$
So, the square of the denominator is given by  $$|c_{++}|^4 + |c_{++}|^2|c_{+-}|^2 + |c_{++}|^2 |c_{-+}|^2 + |c_{+-}|^2 |c_{-+}|^2.$$ On the other hand, the 
$AB$-disentanglement equation  gives us $$|c_{++}|^2( 1- |c_{++}|^2 - |c_{+-}|^2 - |c_{-+}|^2) = |c_{+-}|^2 |c_{-+}|^2.$$ Hence, denominator equals to  $|c_{++}|$ and $\lambda_+ \mu_- = |c_{+-}|.$

Now consider the case in that one of  denominators in (\ref{PR1}), (\ref{PR2}) equals to zero, say $c_{++}=0, c_{-+}=0.$ In this case amplitude factorization can be shown directly. We just set $\lambda_- = |c_{--}|, \lambda_+ = |c_{+-}|, \mu_-=1, \mu_+=0.$

\medskip

 Since some states belonging to ${\cal E}$ can be amplitude factorisable, embedding ${\cal E}_{AB} \to  {\cal E}$ is not surjection.

Gnerally for a state $|\psi\rangle,$  we set $\langle A \rangle_\psi  = \langle \psi| A |\psi \rangle, \langle B \rangle_\psi  = \langle \psi| B |\psi \rangle, \langle A B \rangle_\psi  = \langle \psi|  AB |\psi \rangle.$ Covariance of quantum observables $A$ and $B$ is defined as
\begin{equation}
\label{G4mt}
\rm{cov}(A,B)= \langle   (A - \langle A \rangle_\psi)(B- \langle B \rangle_\psi) \rangle_\psi= 
\langle AB  \rangle_\psi - \langle A \rangle_\psi \langle B \rangle_\psi.
\end{equation}
Observables are called $AB$-uncorrelated in state $|\psi\rangle$ if  $\rm{cov}(A,B) = 0$ or correlation factorization condition holds:
\begin{equation}
\label{ET1}
\langle A B \rangle_\psi = \langle A \rangle_\psi \langle B \rangle_\psi.
\end{equation}
Otherwise they are called correlated. i.e., if $\rm{cov}(A,B)\not=0.$

\medskip
 
{\bf Theorem 3.} {\it A pair of compatible dichotomous observables $A$ and $B$ are $AB$-entangled in  a state $|\psi \rangle$  iff they are correlated: $\rm{cov}(A,B)\not=0,$ i.e.,  }  
\begin{equation}
\label{ET1a}
\langle A B\rangle_\psi \not= \langle A \rangle_\psi \langle B\rangle_\psi
\end{equation} 

\medskip

{\bf Proof.} Set $a=|c_{++}|^2, b= |c_{--}|^2, c= |c_{+-}|^2, d= |c_{-+}|^2.$ Set $x= a+b, y= c+d.$ We remark that 
$x+y=1.$ The latter implies that $x^2 - y^2= x-y.$ Hence, 
$$
\langle A B\rangle_\psi= a+b - c - d=  x-y=  x^2 - y^2 = a^2+b^2- c^2-d^2 + 2ab - 2cd.
$$ 
Then 
$$
\langle A \rangle_\psi \langle B\rangle_\psi= ((a+c) -(d+b))((a+d) - (b+c))=a^2+b^2- c^2-d^2 - 2ab + 2cd 
$$
$$
= x^2 - y^2+ 4ab -4 cd.
$$
Hence, the correlation $\langle A B\rangle_\psi$ is factorisable if and only if $ab=cd,$ i.e., the amplitude factorization condition 
(\ref{E5b}) is satisfied.

\medskip

This proposition supports our invention of entanglement as dependence of  observables. 
Disentanglement corresponds to the classical probabilistic situation of independence of observables given by random variables $A$ and $B.$ For such random variables, independence implies $\langle A B\rangle_P= \langle A \rangle_P \langle B\rangle_P,$ where $P$ is the probability measure (it plays the role of a quantum state). For dichotomous random variables independence is equivalent uncorrelation (see section \ref{dep}). 

It is easy to show that in the quantum theory factorization of correlations does not imply the standard disentanglement, because the amplitude factorization condition can hold for an entangled state. In contrast, by Theorem 3 $AB$-disentanglement is equivalent to correlation factorization for these observables. 

\subsection{State space of arbitrary dimension}

Now we turn to section \ref{eigen}.  Let in superposition (\ref{E2}) all vectors $\vert \psi_{\alpha \beta} \rangle \not=0.$ 
In particular, this presumes that all subspaces ${\cal H}_{AB}(\alpha \beta)$ are nontrivial. 

We rewrite  this superposition as (\ref{E2b}), where 
\begin{equation}
\label{u0}
c_{\alpha \beta} = ||\psi_{\alpha \beta}||
\end{equation}
 and $\vert \alpha \beta \rangle = \vert\psi_{\alpha \beta} \rangle / ||\psi_{\alpha \beta}||.$ Hence, we again can restrict consideration  to the four dimensional case.
Instead of the expansion of a state in the basis of joint eigenvectors of the operators $A$ and $B,$ we consider its expansion 
in its projections on joint eigensubspaces ${\cal H}_{AB} (\alpha \beta),  \alpha \beta= \pm 1.$ We remark that here we do not consider any tensor product structure. 

The violation of $AB$-entanglement condition (\ref{E5}) has the form of equality: 
\begin{equation}
\label{E577}
\Vert  \psi_{+-} \Vert^2 \Vert  \psi_{-+} \Vert^2  = \Vert  \psi_{--}\Vert^2 \Vert  \psi_{++}\Vert^2.
\end{equation}

{\bf Definition 4a.} {\it A state $|\psi \rangle$ is called $AB$ amplitude-factorisable if}
\begin{equation}
\label{Ea1}
||\psi_{\alpha \beta}||= \lambda_\alpha \mu_\beta, \;  \alpha, \beta= \pm,   
\end{equation}
{\it where $\lambda_\alpha, \mu_\beta \geq 0$ and }
\begin{equation}
\label{Ea2}
\lambda_+^2  + \lambda_-^2 =1, \; \mu_+^2 + \mu_-^2= 1.
\end{equation}

\medskip

{\bf Theorem 2a.} {\it For compatible dichotomous observables $A$ and $B,$ a state $|\psi \rangle$ is $AB$-entangled if and only if it is amplitude non-factorisable.}

{\bf Proof.}  $AB$-disentangled is characterized by the equality (\ref{E577}). This equality is equivalent to the amplitude factorization.  The proof is in the line with the proof for the four dimensional 
case. It is evident that amplitude factorization implies equality (\ref{E577}) Now let the latter hold true. We set    
\begin{equation}
\label{PR1a}
\lambda_+=\frac{||\psi_{++}||}{\sqrt{||\psi_{++}||^2+ ||\psi_{-+}||^2}},  \lambda_-=\frac{||\psi_{-+}||}{\sqrt{||\psi_{++}||^2 + ||\psi_{-+}||^2}}.
\end{equation} 
\begin{equation}
\label{PR2a}
\mu_+= \frac{||\psi_{++}||}{\sqrt{||\psi_{++}||^2 + ||\psi_{+-}||^2}},
\mu_-=\frac{||\psi_{-+}||}{\sqrt{||\psi_{++}||^2 + ||\psi_{+-}||^2}}.
\end{equation} 
It is easy to check that these quantities factorize the modules of the coordinates $c_{\alpha, \beta}.$  The cases of zero denominators are analyzed similarly to the four dimensional 
case.

\medskip

As in the the four dimensional case, we obtain  characterization of $AB$-entanglement as observables' correlation:

{\it  Theorem 3 is valid in the general case.}
 
\section{Analog of singlet state for observables entanglement }
\label{EX}

Consider a family of pairs of operators $( A_u,  B_u),$ where $u$ is 
some parameter. Can one find a quantum state  $|\psi \rangle$ that is $A_u B_u$-entangled for all these pairs?   

We assume that, for any $u,$  $[ A_u,  B_u] =0,$ but it may be that, for some pairs $u,v,$
$[ A_u,  A_v] \not =0$ or (and)  $[ B_u,  B_v] \not =0.$ 

Let us consider the tensor  product case ${\cal H}= H \otimes H,$ where $\rm{dim} \; H=2,$ and two operators in $H,  a,  b,$  with the eigenbases $(f_+, f_-)$ and $(g_+, g_-),$ where $ a f_{\pm}=   \pm  f_{\pm},  b g_{\pm}=  \pm   g_{\pm}.$ 
Set $ A=  a \otimes I,  B= I \otimes b.$

Take now any unitary transformation $ u$  in $H$ and consider operators of the form
\begin{equation}
\label{A77}
  a_u =   u  a  u^\star,  b_u =  
 u  b  u^\star,
\end{equation}
which are diagonal w.r.t. to the bases  $(f_{u,+}, f_{u,-})$ and $(g_{u,+}, g_{u,-}),$  
obtained with the unitary transformation $ u$ from the bases $(f_+, f_-)$ and $(g_+, g_-).$

Then consider state $\vert \psi \rangle$ having the form
\begin{equation}
\label{Ltax}
\vert \psi \rangle =  c_{+-} \vert f_+ g_-\rangle + c_{-+} \vert f_- g_+\rangle,\; c_{-+}, c_{-+} \not=0.
\end{equation}
This state is $AB$-entangled. And itpreserves its form under the unitary transformation $U= u\otimes u$ if and only if $c_{+-}= - c_{-+},$ i.e., 
\begin{equation}
\label{A77a}
\vert \psi \rangle =  (\vert f_+ g_-\rangle - \vert f_- g_+\rangle)/ \sqrt{2}= (\vert f_{u,+} g_{u,-}\rangle - 
\vert f_{u,-} g_{u,+}\rangle)/ \sqrt{2}. 
\end{equation}
Hence, this state is $AB$-entangled both for the pair $ A,  B$ and the pair $ A_u= a_u\otimes I , 
 B_u= I \otimes  b_u$ for any unitary transformation $u: H \to H.$

In particular, let us consider the standard singlet state   
\begin{equation}
\label{Ltayz}
\vert \psi_{\rm{singlet}} \rangle =  (\vert f_+ f_-\rangle - \vert f_- f_+\rangle)/\sqrt{2}.  
\end{equation}
This state is $AA^\prime$-entangled for the operators $A=a \otimes I$ and $A^\prime=I \otimes a.$ And it is also 
$A_{u} A_{u}^\prime$-entangled for the operators $A_{u}=a_u \otimes I$ and $A_{u}^\prime=I \otimes a_u.$ 

The singlet state is entangled (in the standard non-factorization sense) independently from the selection of bases, i.e., the operators $A$ and $B.$ However, this is not the case of the operator based entanglement. It is easy to find $u$ such that 
$\vert \psi_{\rm{singlet}} \rangle$ is $AA_{u}^\prime$-disentangled, i.e., amplitude factorisable (cf. discussion in section \ref{4d}).
Let $$
\vert f_-\rangle = u_{-+}  |f_{u,+}\rangle + u_{--}  |f_{u,-}\rangle, \;  
\vert f_+\rangle = u_{++}  |f_{u,+}\rangle + u_{+-}  |f_{u,-}\rangle,
$$
where $u_{-+} \bar{u}_{++} + u_{--} \bar{u}_{+-}=0,\; |u_{-+}|^2 +   |u_{--}|^2 =1, \; |u_{++}|^2 +   |u_{+-}|^2 =1.$
Then 
$$
\vert \psi_{\rm{singlet}} \rangle = (u_{-+}  |f_{+} f_{u,+}\rangle + u_{--}  |f_{+} f_{u,-}\rangle +
u_{++}  |f_{-} f_{u,+}\rangle -  u_{+-}  |f_{-} f_{u,-}\rangle) /\sqrt{2}.
$$
Now select $u_{-+} = u_{+-}= u_{--}= - u_{++}= 1/\sqrt{2}.$ Then  
$$
\vert \psi_{\rm{singlet}} \rangle = (|f_{+} f_{u,+}\rangle +  |f_{+} f_{u,-}\rangle +
 |f_{-} f_{u,+}\rangle - |f_{-} f_{u,-}\rangle) /2.
$$
This state is amplitude factorisable (but not tensor product factorisable) and hence  $AA_{u}^\prime$-disentangled.

Starting with the conditional probabilistic definition of $AB$-entanglement, we analyzed its representation  in the complex HIlbert space formalism and its interrelation (in the tensor product case) with the standard entanglement. 

For compatible dichotomous observables, $AB$-entanglement is equivalent to amplitude non-factorization, where amplitudes are state's projections on joint eigensubspaces of the operators. This characterization leads to the equivalence of $AB$-entanglement to correlation of observables. With such construction, we can proceed in an arbitrary Hilbert space, i.e., without referring to the tesnor product structure.

\section{$AB$-entanglement vs. dependence of classical random variables}
\label{dep}

We recall that in classical probability theory discrete random variables $A$ and $B$ are independent iff their joint probability distribution (JPD) is factorisable, i.e., for all $\alpha, \beta,$
\begin{equation}
\label{a1}
P(A= \alpha, B= \beta)= P(A= \alpha) P(B= \beta).
\end{equation}
This definition is equivalent to reduction of conditional probabilities to ``absolute probabilities'':
\begin{equation}
\label{a2}
P(B= \beta|A= \alpha)= P(B= \beta),\; \mbox{if}\; P(A= \alpha)\not=0,
\end{equation}
\begin{equation}
\label{a3}
P(A= \alpha |B= \beta)= P(A= \alpha), \; \mbox{if}\; P(B= \beta)\not=0.
\end{equation}

Thus, for non-zero probabilities $P(A= \alpha)$ and $P(B= \beta)$ independence -- probability factorization (\ref{a1}), is equivalent to
conditional independence (\ref{a2}), (\ref{a3}).

In quantum theory $AB$-disentanglement means that there exists an outcome $B=\beta$ such that 
\begin{equation}
\label{a4}
P(B= \beta|A= \alpha)= P(B= \beta|A= \alpha^\prime)
\end{equation}
for all $\alpha, \alpha^\prime$ (in the case of $ P(A= \alpha)\not=0, \alpha \in X_a).$

In classical probability, for  random variables $A$ and $B,$ equality (\ref{a4}) is equivalent to equality  (\ref{a2}) 
 -- for the concrete outcome $B=\beta$ (for non-zero probabilities for $A$-outcomes). We prove this to illustrate the role of one of the basic laws of classical probability theory -- the {\it formula of total probability} (FTP):
\begin{equation}
\label{a5}
P(B= \beta)= \sum_\alpha P(B= \beta|A= \alpha)P(A= \alpha).
\end{equation}
Denote the quantity determined by (\ref{a4}) as $P(B= \beta|A).$ We want to show that it coincides with $P(B= \beta).$ We put 
 $P(B= \beta|A)$ in (\ref{a5}):
\begin{equation}
\label{a6}
P(B= \beta)= P(B= \beta|A)) \sum_\alpha  P(A= \alpha)= P(B= \beta|A)).
\end{equation}
Thus, we obtain (\ref{a2}) and hence (\ref{a3}) as well as (\ref{a1}) -- for the concrete value $\beta.$ 

Generally in quantum probability theory FTP is violated, classical FTP is additively perturbed by  the interference term:
\begin{equation}
\label{a7}
P(B= \beta)= \sum_{\alpha, \alpha^\prime} \langle \psi| E_A(\alpha) E_B(\beta) E_A(\alpha^\prime) | \psi \rangle
\end{equation}
$$
= \sum_\alpha P(B= \beta|A= \alpha)P(A= \alpha) + \delta_{AB}(\beta),
$$
where 
\begin{equation}
\label{aa7}
\delta_{AB}(\beta)=\sum_{\alpha \not=\alpha^\prime} \langle \psi| E_A(\alpha) E_B(\beta) E_A(\alpha^\prime) | \psi \rangle.
\end{equation}
In RHS, the first summand corresponds to classical FTP and the second one is the interference term. \footnote{This quantum FTP, FTP with the interference, term was analyzed in very detail in the series of works of one of the authors, e.g., \cite{KHR5}. This FTP plays the basic role in the V\"axj\"o interpretation of QM  \cite{Vaxjo2004,KHR5}.}

By trying to perform the above calculations we obtain 
\begin{equation}
\label{a8}
P(B= \beta) - P(B= \beta|A)) = \delta_{AB}(\beta). 
\end{equation}
Thus,  generally the notion of $AB$-disentanglement differs from the classical notion of independence, formulated as (\ref{a2}). 
 What is about the basic definition of classical independence via probability factorization, see (\ref{a1})?

In quantum probability sequential JPD is defined as
\begin{equation}
\label{G4m}
 P_{AB}(A=\alpha, B= \beta) \equiv P(A=\alpha) P(B=\beta|A=\alpha).
\end{equation}
In operator terms 
\begin{equation}
\label{JPD}
P_{AB}(A=\alpha, B= \beta) = \langle  \psi| E_A(\alpha) E_B(\beta) E_A(\alpha)|  \psi \rangle= ||E_B(\beta) E_A(\alpha)  \psi||^2 .
\end{equation}
For non-commuting observables,  generally 
\begin{equation} 
\label{JPD1}
P_{AB}(A=\alpha, B= \beta)= ||E_B(\beta) E_A(\alpha)  \psi||^2 \not=
\end{equation}
$$
 ||E_A(\alpha)  E_B(\beta) \psi||^2 =P_{BA}( B= \beta, A=\alpha).
$$

{\bf Definition 5.} {\it Observables $A$ and $B$ are called  $AB$-independent w.r.t. state $|\psi\rangle,$ if their sequential JPD is factorisable, i.e., for all pairs $\alpha, \beta,$} 
\begin{equation}
\label{G4ma}
P_{AB}(A= \alpha, B= \beta)= P(A= \alpha) P(B= \beta).
\end{equation}

Otherwise observables are called $AB$-dependent in $|\psi\rangle.$ Since JPD is the sequential JPD, $AB$- and $BA$-(in)dependence are not equivalent. (For commuting observables, $AB$- and $BA$-(in)dependence are equivalent). 

It is easy to check that two sorts of $AB$-independence, i.e., via (\ref{a2}), (\ref{a3}) and (\ref{G4ma}) are equivalent even for quantum probability (for non-zero probabilities of outcomes).
They can be used as another approach to the notion of $AB$-conditional probability (dis)entanglement. 

The most interesting for application is entanglement for compatible observables, $[A, B]=0.$ Here quantum FTP (\ref{a7}) is reduced to classical FTP (\ref{a6}), since the interference  term (\ref{aa7}) equals to zero.

{\bf Proposition 5.} {\it For compatible observables $A$ and $B$ and state $|\psi\rangle,$  the notions of $AB$-entanglement and $AB$-dependence are equivalent.}

{\bf Proof.} 1). Let quantum observables be dependent. Then there exists outcomes $\beta, \alpha$ such that
\begin{equation}
\label{G4mca}
P_{AB}(A= \alpha, B= \beta) \not= P(A= \alpha) P(B= \beta).
\end{equation}
In particular. this implies that $P(A= \alpha)\not=0,  P(B= \beta)\not=0,$ since for commuting operators
$P_{AB}(A= \alpha, B= \beta) = ||E_B(\beta) E_A(\alpha) \psi||^2 \leq || E_A(\alpha) \psi||^2= P(A= \alpha)$ and in the same way
$P_{AB}(A= \alpha, B= \beta) \leq P(B= \beta).$
Thus, we have
\begin{equation}
\label{G4mcab}
P(B= \beta|A= \alpha) \not= P(B= \beta).
\end{equation}
Now suppose that $|\psi\rangle$ is not $AB$-entangled. Then there exists an outcome $\beta$ such that  the equality
$P(B= \beta|A= \alpha)= P(B= \beta|A= \alpha^\prime)$ 
holds for all pairs $\alpha, \alpha^\prime.$ Denote the quantity determined by this constraint by $P(B= \beta|A).$ Now we apply FTP, and obtain that 
$P(B= \beta)= P(B= \beta|A).$ But this contradicts (\ref{G4mcab}).

2). Let quantum observables be $AB$-entangled. Then, for each $\beta,$  $P(B= \beta|A= \alpha)\not= P(B= \beta|A= \alpha^\prime)$ for some $\alpha, \alpha^\prime.$
Suppose that  quantum observables are independent, i.e., 
$P_{AB}(A= \gamma, B= \beta) = P(A= \gamma) P(B= \beta)$ for all $\gamma, \beta.$ This implies that 
$P(B= \beta|A= \gamma) = P(B= \beta),$ and hence $P(B= \beta|A= \alpha)= P(B= \beta|A= \alpha^\prime)$ for $\alpha, \alpha^\prime$ considered above.

\medskip

We remark that compatible observables $A$ and $B$ can represented by random variables $\xi_A$ and $\xi_B$ on
classical probability space. In this case dependence of observables corresponds to dependence of random variables. By Proposition 5
$AB$-entanglement also corresponds to dependence of random variables. We formulate this coupling as

{\bf Proposition 6.} {\it For compatible observables,  $AB$-entanglement can be treated as the complex Hilbert space representation of dependence of classical random variables. }

Now we compare the notion of uncorrelated classical random variables and quantum observables in the dichotomous 
case, i.e., $A, B=  \ pm 1.$   We recall the following result.

{\bf Proposition.} {\it Dichotomous random variables are independent iff they are uncorrelated.}

This is classical probabilistic counterpart part of Theorem 3.

\section{Concluding remarks}

In this article we extended the conditional probability approach to entanglement initiated in \cite{JFA} for probability one 
predictions. This approach can be considered as the probabilistic formalization of the original Schr\"odinger's viewpoint on entanglement 
\cite{SCHE,SCHE1}. We recall that he considered entanglement as apart of complex Hilbert space machinery for prediction of outcomes of observables.\footnote{``It $(\psi$-function) is now the means for predicting probability of measurement results. In it is embodied the momentarily-attained sum of theoretically based future expectation, somewhat as laid down in a catalog.		It is the relation- and -determinacy-bridge between measurements and measurements ...'' \cite{SCHE,SCHE1}.} 
And he treated the EPR argument \cite{EPR} in this way. Hence, for Schr\"odinger, entanglement is related to a pair of observables $A$ and $B,$ i.e., a state is entangled w.r.t. to a selected pair, this is $AB$-entanglement. This original Schr\"odinger's treatment of entanglement as observables entanglement was deformed into state-entanglement and highlighting the role of the tensor product structure
(with just a few exceptions, see \cite{Z1,Z2}).

In this article entanglement is defined as dependence of observables $A$ and $B.$ This viewpoint on entanglement was reflected in 
Schr\"odinger's ``Beschränkung"' by using for it the meaning ``restriction''. Of course, it would be better if Schr\"odinger 
would from the very beginning use the word ``dependence'' and associate dependence of classical 
random variables with $AB$-entanglement, as is done in this article. 

\section*{acknowledgments} 

One of the authors (AK) was supported by COST EU-network DYNALIFE, Information, Coding, and Biological Function:
the Dynamics of Life. 

\section*{Appendix: Classical probability} 

Classical probability theory was mathematically formalized in set and measure theoretical framework  by  Kolmogorov in 1933 \cite{K}. 

Let $\Omega$ be a set of any origin, its points are called elementary events.  Consider  a collection of subsets ${\cal F}$ 
of $\Omega$ forming Boolean $\sigma$-algebra, i.e., it is closed w.r.t. countable unions and intersections and the operation of complement. If $\Omega$ is finite, then ${\cal F}$ is collection of all its subsets.  Let $P$ be a probability measure on ${\cal F}.$

The triple ${\cal P}=(\Omega, {\cal F}, P)$ is called probability space.

A random variables  is map $A: \Omega \to \mathbb{R}$ having some special  property  -- measurability.
We consider only random variables with the discrete range of values.  For $A$ and its value $x,$ set 
  $\Omega_{A=x}=\{s \in \Omega: A(s)= x\}$ and  define the probability  distribution of $A$ as 
	$p_A(x)= P(\Omega_{A=x}).$ In the discrete case measurability means that all sets of the form $\Omega_{A=x}$ belong to 
	${\cal F}.$

\end{document}